\documentclass[prd,12pt,aps,superscriptaddress,preprintnumbers,showpacs,tightenlines,floatfix,nofootinbib,amssymb]{revtex4}
\usepackage{epsfig}

\newcommand{\eff}{{\mathrm{eff}}}

\newcommand{\mon}{{\mathrm{mon}}}
\def\absc{(1-\vert c(s,\mu)\vert^2)^{1/2}}
\newcommand{\beq}{\begin{equation}}
\newcommand{\eeq}{\end{equation}}
\newcommand{\beqn}{\begin{eqnarray}}
\newcommand{\eeqn}{\end{eqnarray}}
\newcommand{\bea}[1]{\beq\begin{array}{#1}}
\newcommand{\eea}{\end{array}\eeq}
\newcommand{\eq}[1]{(\ref{#1})}
\newcommand{\dD}{{\cal D}}
\newcommand{\cD}{{\cal D}}
\newcommand{\cH}{{\cal H}}
\newcommand{\cF}{{\cal F}}
\newcommand{\Erf}{{\mathrm{Erf}}}
\newcommand{\const}{{\mathrm{const}}}
\newcommand{\LL}{{I\!\! L}}

\newcommand{\intinf}{\int\limits^\infty_{-\infty}}

\newcommand{\dual}[1]{{}^{*}{#1}}

\newcommand{\sign}{\mathop{\rm sign}}

\newcommand{\dd}{{\mathrm d}}
\newcommand{\Z}{Z\!\!\! Z}

\newcommand{\diff}{\partial}
\newcommand{\cC}{{\cal C}}
\newcommand{\cZ}{{\cal Z}}
\newcommand{\cS}{{\cal S}}
\newcommand{\cG}{{\cal G}}

\newcommand{\Kanazawa}{\affiliation{Institute for Theoretical Physics,
Kanazawa University, Kanazawa 920-1192, Japan}}

\newcommand{\ITEP}{\affiliation{Institute of Theoretical and
Experimental Physics, B.Cheremushkinskaya 25, Moscow, 117259, Russia}}

\begin{document}

\title{Determination of monopole condensate from monopole action \newline in quenched SU(2) QCD}

\author{M.N.~Chernodub}\Kanazawa\ITEP
\author{Katsuya~Ishiguro}\Kanazawa
\author{Tsuneo~Suzuki}\Kanazawa

\preprint{KANAZAWA/2003-20}
\preprint{ITEP-LAT-2003-15}

\begin{abstract}
We study the effective monopole action obtained in the Maximal Abelian projection
of quenched SU(2) lattice QCD. We determine the quadratic part of the lattice action
using analytical blocking from continuum dual superconductor model to the lattice model.
The leading contribution to the quadratic action depends explicitly on value of the
monopole condensate. We show that the analytical monopole action matches the numerically
obtained action in quenched SU(2) QCD with a good accuracy. The comparison of numerical and
analytical results gives us the value of the monopole condensate in quenched SU(2) QCD,
$\eta = 243(42)$~MeV.
\end{abstract}


\date{August 18, 2003}

\maketitle

\section{Introduction}
\label{one}

The dual superconductor mechanism~\cite{DualSuperconductor} is one of the most promising
mechanisms invented to explain the confinement of color in non--Abelian gauge theories.
The basic element of this mechanism is the existence of specific field
configurations -- called Abelian monopoles -- in the QCD vacuum.
The monopoles are identified with the help of the Abelian projection method~\cite{AbelianProjections},
which uses the partial gauge fixing of the SU(N) gauge symmetry up to an
Abelian subgroup. The Abelian monopoles appear naturally in the Abelian gauge as a result of
the compactness of the residual Abelian group.

Various numerical simulations indicate that the Abelian monopoles may be responsible
for the confinement of quarks (for a review, see, $e.g.$, Ref.~\cite{Reviews}).
The Abelian monopoles provide a dominant contribution to the tension of the fundamental
chromoelectric string~\cite{AbelianDominance,shiba:string,koma:string}.
In Ref.~\cite{MonopoleCondensation} it was
qualitatively shown that the monopole condensate is formed in the low temperature (confinement)
phase and it disappears in the high temperature (deconfinement) phase. The energy profile of the
chromoelectric string as well as the field distribution inside it can be described with a
good accuracy by the dual superconductor model~\cite{string:profile,koma:string,ref:koma:suzuki}

There were various attempts to determine the dual lagrangian and the values of its
couplings~\cite{ref:suzuki:maedan,baker,various:attempts,ref:bali:string,singh,string:profile,ref:koma:suzuki}.
A simplest version of the dual superconductor model for SU(2) gauge theory contains three independent parameters:
the mass of the monopole, $M_\Phi$, the monopole charge, $g$, and the value of the monopole
condensate, $\eta$. In Ref.~\cite{string:profile} the SU(2) string profile was compared
with the classical string solution of the dual superconductor in continuum and the mass
of the dual gauge boson, $M_B=g \eta$, and the monopole mass were shown to be equal\footnote{In this paper
we quote the first set of parameters of Ref.~\cite{string:profile} which is self--consistent.},
$M_B \approx M_\Phi \approx 1.3$~GeV. These values are close to the results of other groups.

The value of the monopole condensate derived from
the chromoelectric string analysis of Ref.~\cite{string:profile} is $\eta=194(19)$~MeV. In
this paper we determine the value of the monopole condensate from the effective monopole
action obtained in the numerical simulations of quenched SU(2) QCD. Our strategy is the following.
We relate the {\it lattice} monopole model on the lattice with the
{\it continuum} dual superconductor model using the approach of blocking of the continuum
variables to the lattice proposed in Ref.~\cite{BlockingOfFields}.
Generally, this method allows to construct perfect lattice actions and operators in
various field theories. In particular, this method was used
in Ref.~\cite{ref:ishiguro:suzuki:chernodub} for the quenched SU(2) QCD
at high temperatures to study the dynamics of the static monopoles.
The lattice monopole action obtained with the help of such a blocking depends on the
parameters of the original continuum model. The comparison of the analytical form of
the lattice monopole action with the corresponding numerical results allows in general to
fix the parameters of the continuum model. In this paper we concentrate on the determination
of the monopole condensate in the quenched SU(2) QCD in the Maximal Abelian projection~\cite{kronfeld}.

The plan of the paper is the following. In Section~\ref{sec:blocking} we propose the method
of blocking from continuum to the lattice of the monopole currents in four dimensional space--time.
We compute the quadratic part of the monopole action analytically in Section~\ref{sec:action:analytically},
while the numerical computation is done in Section~\ref{sec:action:numerically}.
In Section~\ref{sec:comparison} we compare the numerical data with analytically calculated
action and fix the value of the monopole condensate. Our conclusions are presented in the last Section.

\section{Blocking from continuum in four dimensions}
\label{sec:blocking}

\begin{figure}[ht]
\begin{center}
\begin{tabular}{cc}
\includegraphics[angle=-0,scale=1.5,clip=true]{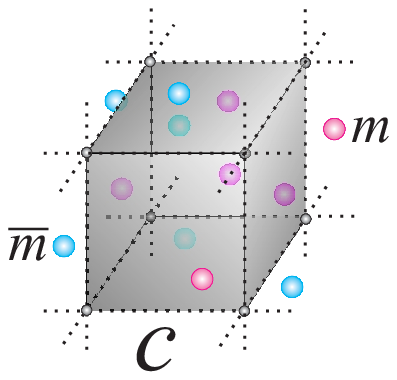} &
\hspace{20mm}  \includegraphics[angle=-0,scale=1.5,clip=true]{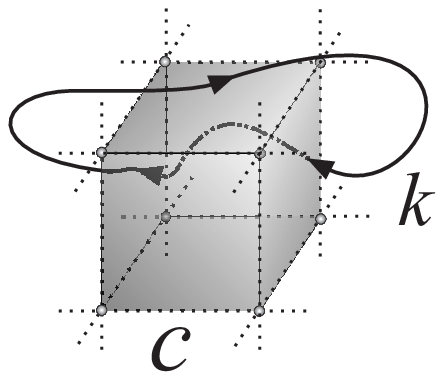}\\
\hspace{-8mm} (a)  & \hspace{12mm}  (b)
\end{tabular}
\end{center}
\vskip -3mm
\caption{Blocking of the continuum monopoles to the lattice in (a) $3D$ and (b) $4D$.
In $3D$ the charge corresponding to the lattice cube $\cC$ is given by the total magnetic charge
of the continuum monopoles inside this cube. In $4D$ the charge is proportional to the
linking number of the monopole trajectory, $k$, with the surface of the three--dimensional cube $\cC$.}
\label{fig:linking}
\end{figure}

The method of blocking of continuum variables to the lattice~\cite{BlockingOfFields,ref:ishiguro:suzuki:chernodub}
constructs the lattice model (at given finite lattice spacing $b$) starting from a model
in continuum. The essence of this method is simple. Consider, for example, the blocking of the
topological variables, such as the monopole charge in three space--time dimensions~\cite{ref:ishiguro:suzuki:chernodub}.
In $3D$ the monopoles are instantons characterized by their positions and the magnetic charges.
Suppose, that the dynamics of these monopole charges in continuum is described by a Coulomb gas model
with two parameters, the fugacity, $\zeta$, and the monopole charge, $g$. Let us superimpose a
cubic lattice with the lattice spacing $b$ on a particular configuration of the monopoles. Each of the
lattice $3D$ cells can be characterized by an integer magnetic charge it contains. Thus we can relate
the continuum configuration of the monopoles to the lattice configuration characterized by
magnetic charge inside each cell (see Figure~\ref{fig:linking} for an illustration).
Next step is to construct a "lattice quantity" (for example, the absolute value
of the magnetic charge inside a $3D$ cell) and calculate analytically the average of this quantity over all
configurations of the continuum monopoles. The value of this averaged quantity would depend on the
size of the cell, $b$, and on the parameters of the continuum model ($i.e.$, on $\zeta$ and $g$).
Similarly, one can study numerically the same quantity in a purely lattice model ($i.e.$,
in the dimensionally reduced quenched SU(2) QCD as in Ref.~\cite{ref:ishiguro:suzuki:chernodub}), and relate both numerical
and analytical results for the density with each other. Since the averaged density depends on the
scale $b$, the fitting of the numerical results to the analytically obtained formula gives an
information about the parameters of the continuum model, $\zeta$ and $g$. The fitting also
provides an information about the self--consistency of this approach, or, in other words, about the
validity of the description of the lattice quantities by the continuum model.

Therefore this method allows us to describe the lattice observables by the continuum model.
In Ref.~\cite{ref:ishiguro:suzuki:chernodub} the blocking was performed for the monopoles in $3D$ which are
the instanton--like objects. Below we generalize this approach to the four dimensional case.

The partition function of the dual superconductor can be described in terms of the
monopole trajectories as follows:
\beqn
\cZ_{mon} = \int\hspace{-4.3mm}\Sigma \dD k \int \dD B \, \exp\Bigl\{
- \int \dd^4 x\, \Bigl[\frac{1}{4 g^2} F^2_{\mu\nu} + i k_\mu(x) B_\mu(x)\Bigr] - S_{int}(k)\Bigr\}\,,
\label{eq:Zmon}
\eeqn
where $F_{\mu\nu} = \partial_\mu B_\nu - \partial_\nu B_\mu$ is the field stress tensor of
the dual gauge field $B_\mu$, and $S_{int}(k)$ is the action of the closed monopole currents
$k$,
\beqn
k_{\mu}(x) = \oint \dd \tau\, \frac{\partial {\tilde x}_\mu(\tau)}{\partial \tau} \,
\delta^{(4)} [x -\tilde x(\tau)]\,.
\eeqn
Here the $4D$ vector function ${\tilde x_\mu}(\tau)$ defines the trajectory of the monopole current.
In Eq.~\eq{eq:Zmon} the integration is carried out over the dual gauge fields and over all possible monopole
trajectories (the sum over disconnected parts of the monopole trajectories is also implicitly assumed).

The action in Eq.~\eq{eq:Zmon} contains three parts: the kinetic term for the dual gauge field, the interaction
of the dual gauge field with the monopole current and the self--interaction of the monopole currents.
The integration over the monopole trajectories gives the lagrangian of the dual Abelian Higgs
model~\cite{ref:suzuki:maedan}:
\beqn
\cZ_{mon} \propto \cZ_{DAHM} = \int \dD \Phi \int \dD B \, \exp\Bigl\{
- \int \dd^4 x \, \Bigl[\frac{1}{4 g^2} F^2_{\mu\nu} + \frac{1}{2} |(\partial_\mu
+ i B_\mu)\Phi|^2 + V(\Phi)\Bigr\}\,,
\label{eq:ZAHM}
\eeqn
where $\Phi$ is the complex monopole field. The self--interactions of the monopole trajectories
described by the action $S_{int}$ in Eq.~\eq{eq:Zmon} lead to the self--interaction of the
monopole field $\Phi$ described by the potential term $V(\Phi)$ in Eq.~\eq{eq:ZAHM}.

Now let us embed the hypercubic lattice with the lattice spacing $b$ into the continuum space.
The three-dimensional cubes are defined as follows:
\beqn
C_{s,\mu} = \Biggl\{b \Bigl(s_\nu - \frac{1}{2}\Bigr) \leq x_\nu \leq
b \Bigl(s_\nu + \frac{1}{2}\Bigr)\,\, \mbox{for}\,\, \nu\neq\mu\,; \,\, \mbox{and} \,\,
x_\mu = b s_\mu \Biggr\}\,,
\label{eq:C}
\eeqn
where $s_\nu$ is the dimensionless lattice coordinate of the
lattice cube $C_{s,\mu}$ and $x_\nu$ is the continuum coordinate. The direction
of the $3D$ cube in the $4D$ space is defined by the Lorentz index $\mu$.

As in the three-dimensional example described above, let us consider a configuration
of the monopole currents superimposed on the lattice~\eq{eq:C}.
The monopole charge $K_C$ inside the lattice cube $C_{s,\mu}$ is equal to the total charge
of the continuum monopoles, $k$, which pass through this cube. Geometrically, the total
monopole corresponds to the linking number between the cube $C$ and the monopole
trajectories, $k$ (an illustration is presented in Fig.~\ref{fig:linking}). The mutual orientation
of the cube and the monopole trajectory is
obviously important. The corresponding mathematical expression for the monopole charge
$K_C$ inside the cube $C$ is a generalization of the Gauss linking number to the four
dimensional space--time:
\beqn
 K_C(k) \equiv \LL(\partial C,k) & = & \frac{1}{2} \int \!\dd^4 x \int \!
 \dd^4 y \, \epsilon_{\mu \nu \alpha \beta} \, \Sigma^{\partial C}_{\mu\nu}(x)\,
 k_{\alpha}(y)  \, \diff_{\beta} \cD^{(4)}(x - y) \nonumber \\
 & = & \frac{1}{4 \pi^2}
 \int \dd^4 x \int \dd^4 y\, \epsilon_{\mu\nu\alpha\beta}\,
 \Sigma^{\partial C}_{\mu\nu}(x)\, k_{\alpha}(y)\,\frac{{(x-y)}_{\beta}
 }{{|x-y|}^4}\,.
\label{eq:Link4D}
\eeqn
Here the function $\Sigma^{\partial C}_{\mu\nu}(x)$ is the two--dimensional $\delta$--function
representing the boundary $\partial C$ of the cube $C$. In general form it can be written as follows:
\beqn
\Sigma_{\alpha \beta}(x) = \int_{\Sigma} \dd \tau_1 \dd \tau_2 \,
\frac{x_{[\alpha,}(\vec \tau)}{\partial \tau_a} \frac{x_{\beta]}(\vec \tau)}{\partial \tau_b}
\, \delta^{(4)} [x -\tilde x(\vec \tau)]\,,
\eeqn
where the four dimensional vector $\tilde x(\vec \tau)$ parameterizes the position of the
two--dimensional surface $\Sigma$. The function $\cD^{(4)}$ in Eq.~\eq{eq:Link4D} is the inverse
Laplacian in four dimensions, $\partial^2_\mu \cD^{(4)}(x) = \delta^{(4)}(x)$.
It is obvious that the lattice currents $K_{s,\mu}$ are closed
\beqn
\partial' K = 0\,,
\label{eq:closeness}
\eeqn due to the conservation of the continuum monopole charge, $\partial_\mu k_\mu =0$.
In Eq.~\eq{eq:closeness} the symbol $\partial'$ denotes the backward derivative on the lattice.
We present a proof of Eq.~\eq{eq:closeness} in Appendix~\ref{app:closeness}.

Let us rewrite the dual superconductor model~\eq{eq:ZAHM} in terms of the lattice currents $K_C$,
Eq.~\eq{eq:Link4D}. To this end we insert the unity,
\beqn
1 = \sum\limits_{K_C\in \Z} \, \prod\limits_C\delta\Bigl( K_C - \LL(\partial C,k)\Bigr)\,,
\label{eq:unity:1}
\eeqn
into the partition function \eq{eq:Zmon} (here $\delta$ represents the Kronecker symbol).
Then we integrate the continuum degrees of freedom, $k_\mu$ and $B_\mu$, getting the
partition function in terms of the lattice charges $K_C$. The simplest way to do so is to represent
the product of the Kronecker symbols in Eq.~\eq{eq:unity:1} in terms of the integrals,
\beqn
1 = \sum\limits_{K_C\in \Z} \, \Bigl[\prod_C \intinf \dd \theta_C\Bigr]\,
\exp\Bigl\{i \sum_C \theta_C \, K_C - i \int \dd^4 x \, k_\mu(x) {\tilde B}_\mu(\theta;x)\Bigr\}\,,
\label{eq:unity:2}
\eeqn
where
\beqn
{\tilde B}_\mu(\theta;x) = \frac{1}{2} \int \! \dd^4 y \, \epsilon_{\mu \nu \alpha \beta} \,
\diff_{\nu} \cD^{(4)}(x - y) \, \sum\limits_C \theta_C \, \Sigma^{\partial C}_{\alpha \beta}(y)\,.
\label{eq:tildeB}
\eeqn
To derive Eqs.~(\ref{eq:unity:2},\ref{eq:tildeB}) from Eq.~\eq{eq:unity:1} we
used relation~\eq{eq:Link4D}.

Substituting Eq.~\eq{eq:unity:2} into Eq.~\eq{eq:Zmon} we get:
\beqn
\cZ_{mon} & = & \int\hspace{-4.3mm}\Sigma \dD k \int \dD B \, \sum\limits_{K_C\in \Z} \,
\Bigl[\prod_C \intinf \dd \theta_C\Bigr]\, \exp\Bigl\{ i \sum_C \theta_C \, K_C \nonumber\\
& & - \int \dd^4 \Bigl[\frac{1}{4 g^2} F^2_{\mu\nu} + i k_\mu(x)
\Bigl(B_\mu(x)+{\tilde B}_\mu(\theta;x)\Bigr)\Bigr] - S_{int}(k)\Bigr\}\,.
\label{eq:Zmon:1}
\eeqn
One can see that the substitution of the unity~\eq{eq:unity:2} effectively shifts the
gauge field in the interaction term with the monopole current, $B_\mu \to B_\mu + {\tilde B}_\mu$.
Therefore the integration over the monopole trajectories, $k_\mu$, in Eq.~\eq{eq:Zmon:1}
is very similar to the integration which relates Eq.~\eq{eq:Zmon} and Eq.~\eq{eq:ZAHM}. Thus, we get:
\beqn
\cZ_{mon} \propto \cZ_{DAHM} & = & \int \dD \Phi \int \dD B \, \sum\limits_{K_C\in \Z} \,
\Bigl[\prod_C \intinf \dd \theta_C\Bigr]\, \exp\Bigl\{ i \sum_C \theta_C \, K_C \nonumber\\
& & - \int \dd^4 x \Bigl[\frac{1}{4 g^2} F^2_{\mu\nu} + \frac{1}{2} \Bigl|\Bigl[ \partial_\mu +
i(B_\mu(x)+{\tilde B}_\mu(\theta;x))\Bigr]\Phi\Bigr|^2 + V(\Phi)\Bigr]\Bigr\}\,.
\label{eq:ZAHM:1}
\eeqn

Summarizing this Section,
we rewrite the continuum dual superconductor model in terms of the lattice monopole currents, $K$:
\beqn
\cZ_{DAHM} = \sum\limits_{K_{x,\mu}\in \Z} \, e^{-S_{mon}(K)}\,,
\label{eq:Zmon:lat}
\eeqn
where the monopole action is defined via the lattice Fourier transformation:
\beqn
e^{-S_{mon}(K)} = \intinf \dD \theta_C \,
\exp\Bigl\{- {\tilde S}(\theta) + i (\theta, K) \Bigr\}\,,
\label{eq:Smon:lat}
\eeqn
and the action of the compact lattice fields $\theta$ is expressed in terms of
the dual Abelian Higgs model in continuum:
\beqn
e^{-{\tilde S}(\theta)} \!= \!\int \!\!\dD \Phi \int \!\!\dD B \, \exp\Bigl\{
- \!\!\int \!\dd^4 x \Bigl[\frac{1}{4 g^2} F^2_{\mu\nu} + \frac{1}{2} \Bigl|\Bigl[ \partial_\mu +
i(B_\mu+{\tilde B}_\mu(\theta)\Bigr]\Phi\Bigr|^2 + V(\Phi)\Bigr]\Bigr\}\,.
\label{eq:Stilde:lat}
\eeqn
Eqs.~(\ref{eq:tildeB},\ref{eq:Zmon:lat},\ref{eq:Smon:lat},\ref{eq:Stilde:lat})
are the main result of this Section.

\section{Quadratic part of monopole action}
\label{sec:action:analytically}

An exact integration over the monopole, $\Phi$, and dual gauge gluon, $B_\mu$, fields in
Eq.~\eq{eq:Stilde:lat} is impossible in a general case. However, in this paper
we are interested in the quadratic part of the monopole action which is dominated
by the contribution of the one dual gluon exchange. Therefore we do not consider the effect
of the fluctuations of the monopole field $\Phi$, which lead to the higher--point
interactions in the effective monopole action\footnote{The fluctuations of the monopole
fields and their effect on the blocked monopole action will be considered in a subsequent
paper.}~\cite{chernodub}. Effectively, the neglect of the quantum fluctuations
of the monopole field corresponds to a mean field approximation with respect to this
field, $\Phi \to \langle \Phi \rangle$.
In this case the AHM action becomes quadratic and Eq.~\eq{eq:Stilde:lat}
can be rewritten as
\beqn
e^{-{\tilde S}(\theta)} \!= \int \!\!\dD B \, \exp\Bigl\{
- \!\!\int \!\dd^4 x \Bigl[\frac{1}{4 g^2} F^2_{\mu\nu} + \frac{\eta^2}{2}
{\Bigl(B_\mu+{\tilde B}_\mu(\theta)\Bigr)}^2\Bigr]\Bigr\}\,,
\label{eq:Stilde:London}
\eeqn
where $\eta = |\langle \Phi \rangle|$ is the monopole condensate.

The Gaussian integration over the dual gauge field can be done explicitly. In momentum
space the effective action (up to an irrelevant additive constant) reads as follows:
\beqn
{\tilde S}(\theta) = \frac{\eta^2}{2} \int \frac{\dd^4 p}{(2 \pi)^4} \,
{\tilde B}_\mu(\theta, p) \, \frac{p^2 \delta_{\mu\nu} - p_\mu p_\nu}{p^2 + M_B^2}\,
{\tilde B}_\mu(\theta, -p)\,,
\label{eq:Stilde:p}
\eeqn
where ${\tilde B}_\mu(\theta, p)$ is related to the field ${\tilde B}_\mu(\theta, x)$,
given in Eq.~\eq{eq:tildeB}, by a continuum Fourier transformation:
\beqn
{\tilde B}_\mu(\theta, p) = \frac{b^3}{p^2}\sum\limits_{s,\alpha}
[p^2\, \delta_{\mu\alpha} Q_\alpha (p b) - p_\mu p_\alpha\, Q_\alpha (p b)] e^{- i b (p,s)} \,
\theta_{s,\alpha}\,,
\label{eq:B:p}
\eeqn
with
\beqn
Q_\mu (x) = \prod\limits_{\nu\neq\mu} \frac{\sin x_\nu /2}{x_\nu/2}\,.
\label{eq:Q}
\eeqn
To get Eq.~\eq{eq:B:p} from Eq.~\eq{eq:tildeB} we notice that
\beqn
\frac{1}{2} \epsilon_{\mu \nu \alpha \beta} \, \Sigma^{\partial C}_{\alpha \beta}(x) =
\partial_{[\mu}, V^C_{\nu]}(x)\,,
\label{eq:epsilon}
\eeqn
where $V^C_\mu$ is the characteristic function of the lattice cell $C_{s,\mu}$. Namely,
the characteristic function of the $3D$ cube with the lattice coordinate $s_\mu$ and
the direction $\alpha$ is
\beqn
V_\mu(C_{s,\alpha},x) = \delta_{\mu,\alpha} \, \delta(x_\alpha - b s_\alpha)
\prod\limits_{\gamma \neq \alpha} \Theta(b (s_\gamma+1/2) - x_\gamma )
\cdot \Theta(x_\gamma - b (s_\gamma-1/2))\,,
\label{eq:V}
\eeqn
where $\Theta(x)$ is the Heaviside function. The Fourier transform of the function~\eq{eq:V}~is
\beqn
V_\mu(C_{x,\alpha},p) = \delta_{\mu,\alpha} \, b^3\, Q_\alpha(p b) \, e^{- i b (p,s)}\,.
\label{eq:V:p}
\eeqn

Substituting Eq.~\eq{eq:B:p} into Eq.~\eq{eq:Stilde:p} and changing the momentum variable, $q=b p$, we
get the following expression for the quadratic action:
\beqn
{\tilde S}(\theta) = \frac{\eta^2 b^2}{2}
\sum_{s,s'} \sum_{\alpha,\alpha'} \theta_{s,\alpha} \cF^{-1}_{ss',\alpha\alpha'} \theta_{s' \alpha}
\theta_{s',\alpha'} \,,
\label{eq:tilde:S}
\eeqn
where
\beqn
\cF^{-1}_{ss',\alpha\alpha'} = \int \frac{\dd^4 q}{(2 \pi)^4}
\frac{q^2 \delta_{\alpha\alpha'} - q_\alpha q_{\alpha'}}{q^2 + \mu^2}
Q_\alpha(q) Q_{\alpha'}(q)\, e^{i q (s' -s)} \,.
\label{eq:F:inverse}
\eeqn
Here we have introduced the dimensionless parameter
\beqn
\mu = M_B \, b\,.
\label{eq:mu}
\eeqn

Next step is to substitute Eq.~\eq{eq:tilde:S} into Eq.~\eq{eq:Smon:lat} and integrate
over the variables $\theta$
to get the quadratic monopole action:
\beqn
S_{\mathrm{mon}}(K) = \sum_{s,s'} \sum_{\alpha,\alpha'}
K_{s,\alpha} \, \cS_{ss',\alpha\alpha'} \, K_{s',\alpha'}\,,\quad
\cS_{ss',\alpha\alpha'} = \frac{1}{2 \, \eta^2 b^2} \cF_{ss',\alpha\alpha'}\,.
\label{eq:S:mon:formal}
\eeqn

We could not find an explicit form for the operator $\cF^{-1}$ and therefore we calculate it
in the $\mu \to \infty$ limit. This limit corresponds to large values of $b$ which are
consistent with the quadratic form of the monopole action~\cite{chernodub}.
The details of the calculation are given in Appendix~\ref{app:cF}, and the result is:
\beqn
\cF^{-1}_{ss',\alpha\alpha'}\!\! & = & \frac{\delta_{\alpha\alpha'}}{4 \pi} \delta_{s_\alpha, s_\alpha'}
\Bigl[ \Gamma(0, t_{UV} \mu^2) \, \cD^{(3)}_\alpha({(\vec s - \vec s')}_\perp) \nonumber\\
& & \qquad \qquad \quad + \frac{2}{\mu} \cG_\alpha({(\vec s - \vec s')}_\perp)
+ \frac{3}{\pi \mu^2} \cH_\alpha({(\vec s - \vec s')}_\perp) \Bigr],
\label{eq:cF:inverse:final} \\
\cD^{(3)}_\alpha (\vec s) & = & \sum_{\stackrel{\mbox{\footnotesize cyclic}}{i,j,k\neq \alpha}}
\Delta_{s_i} \delta_{s_j} \delta_{s_k}\,, \quad
\cG_\alpha (\vec s) = \sum_{\stackrel{\mbox{\footnotesize cyclic}}{i,j,k\neq \alpha}}
\Delta_{s_i} \Delta_{s_j} \delta_{s_k}\,, \quad
\cH_\alpha (\vec s) = \prod_{i \neq \alpha} \Delta_{s_i} \,. \nonumber
\eeqn
Here $\cD^{(3)}_\alpha (\vec s_\perp)$ is the three-dimensional Laplacian acting in
a timeslice perpendicular to the direction $\hat \alpha$, $\delta_s$ is the Kronecker symbol,
$\Delta_s \equiv \cD^{(1)}(s)$ is the one--dimensional Laplacian operator (double derivative)
defined in Eq.~\eq{eq:laplacian}, $\Gamma(a,x)$ is the incomplete gamma function
and $t_{UV}$ is an ultraviolet cutoff.
In Eq.~\eq{eq:cF:inverse:final} exponentially suppressed corrections of
the order $O(e^{-\const \mu})$ are omitted.

Inverting the operator \eq{eq:cF:inverse:final} and expanding it in inverse powers of $\mu$
we get the quadratic operator $\cS$ in the monopole action~\eq{eq:S:mon:formal}:
\beqn
\cS_{ss',\alpha\alpha'} & = & \frac{2 \pi}{\eta^2 b^2\, \Gamma} \cdot \delta_{\alpha\alpha'}
\delta_{s_\alpha,s_\alpha'} \cdot \Bigl[ \cD^{-1}_\alpha - \frac{2}{\mu \Gamma} \,
\cD^{-1}_\alpha \, \cG_\alpha \, \cD^{-1}_\alpha \nonumber\\
& & + \frac{1}{\pi \mu^2 \Gamma^2} \Bigl(4 \pi \, \cD^{-1}_\alpha \, \cG_\alpha \,
\cD^{-1}_\alpha \, \cG_\alpha \, \cD^{-1}_\alpha - 3 \Gamma \cD^{-1}_\alpha
\cH_\alpha \, \cD^{-1}_\alpha\Bigr) + O(\mu^{-3}) \Bigr]_{{(\vec s - \vec s')}_\perp}\,.
\label{eq:cS:expansion}
\eeqn
where $\cD_\alpha \equiv \cD^{(3)}_\alpha$, $\Gamma \equiv \Gamma(0, t_{UV} M^2_B\, b^2)$.
The operator expansion in Eq.~\eq{eq:cS:expansion} is written in a symbolic form.

\section{Monopole action in quenched SU(2) QCD}
\label{sec:action:numerically}

Having determined the action of the blocked monopoles analytically, we are going
to the same in the quenched SU(2) QCD using numerical calculations. We simulate the quenched SU(2)
gluodynamics with the lattice Wilson action, $S(U) = - \frac{\beta}{2} \sum_P {\mathrm{Tr}} U_P$,
where $\beta$ is the coupling constant and $U_P$ is the SU(2) plaquette constructed
from the link fields. All our results are obtained in the Maximal Abelian (MA)
gauge~\cite{kronfeld} which is defined by the maximization of the lattice functional
\beqn
R = \sum_{s,\hat\mu}{\rm Tr}\Big(\sigma_3 \widetilde{U}(s,\mu)
\sigma_3 \widetilde{U}^{\dagger}(s,\mu)\Big)\,,
\label{R}
\eeqn
with respect to the SU(2) gauge transformations
$U(s,\mu) \to \widetilde{U}(s,\mu)=\Omega(s)U(s,\mu)\Omega^\dagger(s+\hat\mu)$.
The local condition of maximization can be written in the continuum limit as the
differential equation $(\partial_{\mu}+igA_{\mu}^3)(A_{\mu}^1-iA_{\mu}^2)=0$.
Both this condition and the functional \eq{R} are invariant under
residual U(1) gauge transformations, $\Omega^{\mathrm{Abel}}(\omega)
= {\mathrm{diag} (e^{i \omega(s)},e^{- i \omega(s)})}$.

After the gauge fixing is done we get the Abelian variables applying
the Abelian projection to the non--Abelian link variables. The Abelian gauge
field is extracted from the $SU(2)$ link variables as follows:
\beqn
 \widetilde{U}(s,\mu) = \left( \begin{array}{cc}
         \absc        & -c^*(s,\mu) \\
                  c(s,\mu) &  \absc
\end{array} \right)
\left( \begin{array}{cc}
u(s,\mu) & 0 \\
0 & u^*(s,\mu)
\end{array} \right),
\label{eq:field:decomposition}
\eeqn
where $u(s,\mu)= \exp(i\theta(s,\mu))$ represents the Abelian link field
and $c(s,\mu)$ corresponds to the charged (off--diagonal) matter fields.
The Abelian field strength $\theta_{\mu\nu}(s)\in(-4\pi,4\pi)$ is defined
on the lattice plaquettes by the Abelian link angle $\theta(s,\mu)\in[-\pi,\pi)$
as follows: $\theta_{\mu\nu}(s)=\theta(s,\mu)+
\theta(s+\hat\mu,\nu)-\theta(s+\hat\nu,\mu)-\theta(s,\nu)$.

To construct the Abelian monopoles we decompose the field strength $\theta_{\mu\nu}(s)$ into two parts,
\beqn
\theta_{\mu\nu}(s)= \bar{\theta}_{\mu\nu}(s) +2\pi m_{\mu\nu}(s)\,,
\label{eq:field:separation}
\eeqn
where $\bar{\theta}_{\mu\nu}(s)\in [-\pi,\pi)$ is interpreted as
the electromagnetic flux through the plaquette
and $m_{\mu\nu}(s)$ can be regarded as a number of the Dirac
strings piercing the plaquette. The elementary ($i.e.$, defined on
the $1^3$ lattice cubes) monopole currents are determined
by the DeGrand-Toussaint\cite{degrand} formula:
\beqn
k_{\mu}(s) & = & \frac{1}{2}\epsilon_{\mu\nu\rho\sigma}
\partial_{\nu}m_{\rho\sigma}(s+\hat{\mu}),
\label{eq:monopole:definition}
\eeqn
where $\partial$ is the forward lattice derivative. The elementary monopole current
is defined on a link of the dual lattice and takes values $0, \pm 1, \pm 2$.
Moreover the elementary monopole current satisfies the conservation condition by
construction,
\beqn
\partial'_{\mu} k_{\mu}(s)=0\,,
\eeqn
where $\partial'$ is the backward derivative on the dual lattice.

Besides the elementary monopoles one can also study the so called
extended mo\-no\-po\-les~\cite{ivanenko}. The extended monopoles are usually used
to define the monopole current on a cube of a large size without getting
artificial lattice corrections of the order of the lattice spacing, $a$.
The $n^3$ extended monopole is defined on a sublattice with the
lattice spacing $b=na$.
The explicit construction of the extended monopoles corresponds to a block spin
transformation of the monopole currents with the scale factor $n$,
\beqn
k_{\mu}^{(n)}(s) = \sum_{i,j,l=0}^{n-1}k_{\mu}(n s+(n-1)\hat{\mu}+i\hat{\nu}
     +j\hat{\rho}+l\hat{\sigma})\,.
\label{eq:blocking}
\eeqn
The spacing $a$ of the original lattice and, consequently, the
artificial lattice corrections (which are of the order of $O(a)$) can
be arbitrarily small while the size of the blocked monopole can be fixed in physical units.
In our studied we have studied $n=2,3,4,6,8$ blocked monopoles on $48^4$ lattice.

Applying consecutively the gauge fixing, the Abelian projection and using
formula~\eq{eq:blocking} one can construct the Abelian monopole ensemble for
any ensemble of the non--Abelian fields of quenched SU(2) QCD. Then using an
inverse Monte-Carlo method one can get the effective monopole action. The
details of this procedure can be found in Refs.~\cite{shiba:condensation,ref:suzuki:shiba,chernodub}.
In our simulations we have used 200 configurations on $48^4$ lattice. The Maximal
Abelian gauge was fixed with the help of the standard iterative procedure.

\begin{figure}[ht]
\includegraphics[angle=-0,scale=0.75,clip=true]{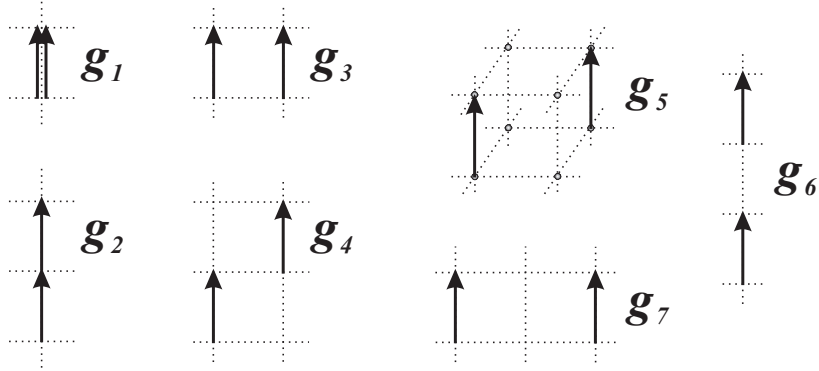}
\caption{The graphic representation of the first seven types of the quadratic interactions
in the lattice monopole action schematized in Table~\ref{tbl:gs}.}
\label{fig:interactions}
\end{figure}

In general, the monopole action, $S^{\mon}_{\eff}$, can be represented as a sum of
the $n$--point ($n \ge 2$) operators $S_i$, Ref.~\cite{shiba:condensation,chernodub}:
\beqn
S[k] = \sum_i g_i S_i [k]\,,
\label{eq:monopole:action}
\eeqn
where $f_i$ are coupling constants. In this paper we adopt only the two--point interactions
of the form $S_i \sim k_{\mu}(s) k_{\mu'}(s')$ which works well at large values of $b$.
Using the inverse Monte-Carlo
method we calculate the monopole action parameterized by 27 couplings $g_i$. The maximal
distance between the interacting currents in this action is $3$ in units of
the blocked lattice spacing $b$. The contributions of higher--distance interactions are
very small. The mutual separations and directions of the monopole currents corresponding
to the couplings $g_i$ are summarized in Table~\ref{tbl:gs}. We visualize the first seven
most essential coupling constants in the monopole action in Figure~\ref{fig:interactions}.
\begin{table}[ht]
\begin{tabular}{|c|l|l||c|l|l|}
\hline
coupling &  distance& $ \ \ \ \ \ \ \ \ $  type $ \ \ \ \ \ \ \ \ $ &
coupling &  distance& $ \ \ \ \ \ \ \ \ $  type \\
\hline
$g_1$    & (0,0,0,0) & $k_\mu(s)$ &
$g_{15}$ & (2,1,1,0) & $k_\mu(s+2\hat{\mu}+\hat{\nu}+\hat{\rho})$ \\
$g_2$    & (1,0,0,0) & $k_\mu(s+\hat{\mu})$ &
$g_{16}$ & (1,2,1,0) & $k_\mu(s+\hat{\mu}+2\hat{\nu}+\hat{\rho})$ \\
$g_3$    & (0,1,0,0) & $k_\mu(s+\hat{\nu})$ &
$g_{17}$ & (0,2,1,1) & $k_\mu(s+2\hat{\nu}+\hat{\rho}+\hat{\sigma})$ \\
$g_4$    & (1,1,0,0) & $k_\mu(s+\hat{\mu}+\hat{\nu})$ &
$g_{18}$ & (2,1,1,1) & $k_\mu(s+2\hat{\mu}+\hat{\nu}+\hat{\rho}+\hat{\sigma})$ \\
$g_5$    & (0,1,1,0) & $k_\mu(s+\hat{\nu}+\hat{\rho})$ &
$g_{19}$ & (1,2,1,1) & $k_\mu(s+\hat{\mu}+2\hat{\nu}+\hat{\rho}+\hat{\sigma})$ \\
$g_6$    & (2,0,0,0) & $k_\mu(s+2\hat{\mu})$ &
$g_{20}$ & (2,2,0,0) & $k_\mu(s+2\hat{\mu}+2\hat{\nu})$ \\
$g_7$    & (0,2,0,0) & $k_\mu(s+2\hat{\nu})$ &
$g_{21}$ & (0,2,2,0) & $k_\mu(s+2\hat{\nu}+2\hat{\rho})$ \\
$g_8$    & (1,1,1,1) & $k_\mu(s+\hat{\mu}+\hat{\nu}+\hat{\rho}+\hat{\sigma})$ &
$g_{22}$ & (3,0,0,0) & $k_\mu(s+3\hat{\mu})$ \\
$g_9$    & (1,1,1,0) & $k_\mu(s+\hat{\mu}+\hat{\nu}+\hat{\rho})$ &
$g_{23}$ & (0,3,0,0) & $k_\mu(s+3\hat{\nu})$ \\
$g_{10}$ & (0,1,1,1) & $k_\mu(s+\hat{\nu}+\hat{\rho}+\hat{\sigma})$ &
$g_{24}$ & (2,2,1,0) & $k_\mu(s+2\hat{\mu}+2\hat{\nu}+\hat{\rho})$ \\
$g_{11}$ & (2,1,0,0) & $k_\mu(s+2\hat{\mu}+\hat{\nu})$ &
$g_{25}$ & (1,2,2,0) & $k_\mu(s+\hat{\mu}+2\hat{\nu}+2\hat{\rho})$ \\
$g_{12}$ & (1,2,0,0) & $k_\mu(s+\hat{\mu}+2\hat{\nu})$ &
$g_{26}$ & (0,2,2,1) & $k_\mu(s+2\hat{\nu}+2\hat{\rho}+\hat{\sigma})$ \\
$g_{13}$ & (0,2,1,0) & $k_\mu(s+2\hat{\nu}+\hat{\rho})$ &
$g_{27}$ & (2,1,1,0) & $k_\rho(s+2\hat{\mu}+2\hat{\nu}+\hat{\rho})$ \\
$g_{14}$ & (2,1,0,0) & $k_\nu(s+2\hat{\mu}+\hat{\nu})$ &
         &           & \\
\hline
\end{tabular}
\caption{The quadratic interactions in the monopole action determined numerically.}
\label{tbl:gs}
\end{table}

The action determined above takes into account all monopole trajectories.
However, a typical monopole configuration in the confinement phase consists of one large
monopole trajectory (percolating cluster) supplemented by a lot of small (ultraviolet) monopole
clusters~\cite{ref:kitahara}. The percolating cluster fills the whole volume of the
lattice and it makes a dominant
contribution to the tension of the chromoelectric string. The properties of the largest
percolating cluster were studied both numerically~\cite{ref:kitahara,ishiguro:entropy,ref:boyko} and
analytically~\cite{ref:zakharov:clusters}. The percolating
cluster is associated with the monopole condensate~\cite{ivanenko:condensate,shiba:condensation}.

\begin{figure}[htb]
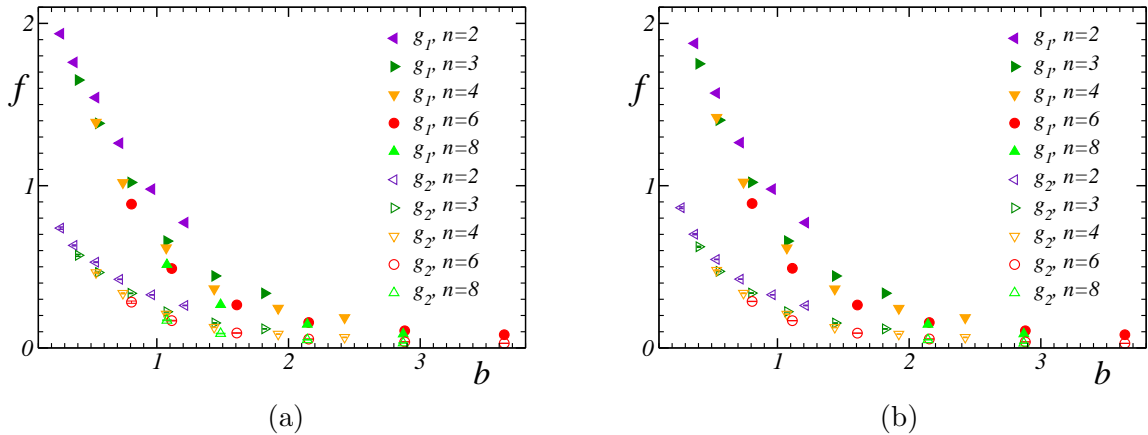

\begin{tabular}{cc}
\includegraphics[angle=-00,scale=0.3,clip=true]{g.all.eps}
&
\hskip 12mm\includegraphics[angle=-0,scale=0.3,clip=true]{g.max.eps}
\\
\hskip 5mm (a) & \hskip 15mm (b)
\end{tabular}
\caption{The couplings $g_1$ and $g_2$ of the monopole action (a) for
all clusters and (b) for the percolating cluster. In this and other figures the error bars are smaller
than the size of symbols and the scale $b$ is shown in units of the string tension.}
\label{fig:g:max:all}
\end{figure}
If our determination of the monopole action is self--consistent then at large scales $b$ the
ultraviolet clusters should not give any contribution neither to the monopole action nor to
the monopole condensate. The correctness of
the first statement for the leading parameter, $g_1$, was confirmed in Ref.~\cite{ishiguro:entropy}.
In Figures~\ref{fig:g:max:all}(a), (b) we show the couplings $g_1$ and $g_2$ for
all clusters and for the percolating cluster. These couplings show an approximate scaling:
they depend only on the product $b=a\cdot n$ and  do not depend on the variables $a$
and $n$ separately when $n\ge 3$ are considered. The larger $b$ the better scaling is.

\begin{figure}[htb]
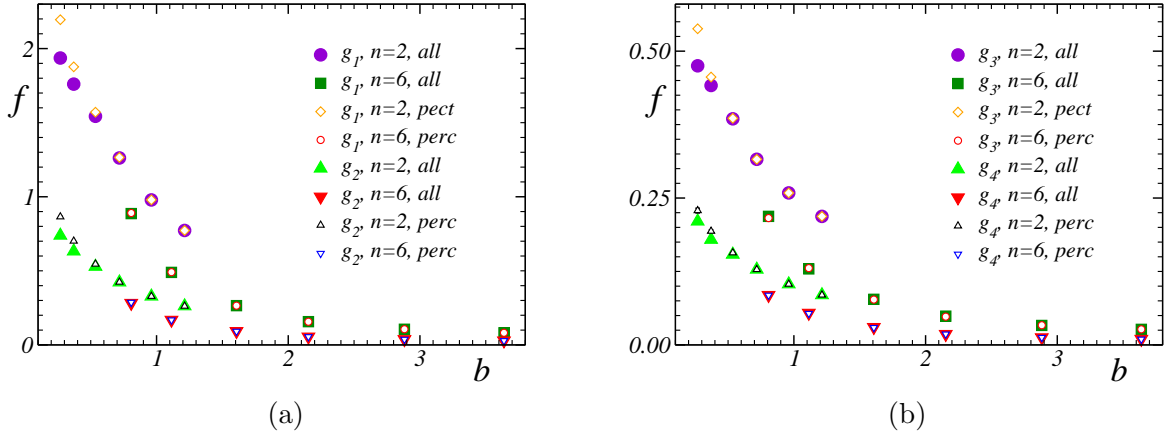

\begin{tabular}{cc}
\includegraphics[angle=-00,scale=0.3,clip=true]{g.compare1.eps}
&
\hskip 12mm\includegraphics[angle=-0,scale=0.3,clip=true]{g.compare2.eps}
\\
\hskip 5mm (a) & \hskip 15mm (b)
\end{tabular}
\caption{The comparison of the couplings (a) $g_{1,2}$ and (b) $g_{3,4}$ computed
for all clusters and for the percolating cluster.}
\label{fig:g:compare}
\end{figure}

The comparison of the couplings computed on all clusters and on the percolating cluster only are
shown in Figures~\ref{fig:g:compare}(a) and (b). Again, one can clearly observe that at
large scales $b$ the coupling constants evaluated on the different types of the monopole ensembles
coincide with each other contrary to the small--$b$ case.

\section{Monopole condensate from monopole action}
\label{sec:comparison}

To get the value of the monopole condensate we have to compare the monopole action
calculated analytically in Section~\ref{sec:action:analytically} with the numerical results
described in Section~\ref{sec:action:numerically}. To this end we first note that due to the closeness of
the monopole currents $K_{x,\mu}$ only the transverse part of the monopole operator~\eq{eq:cS:expansion}
has a sense. Indeed, the shift of the quadratic operator $\cS \to \cS + \alpha \partial \partial'$
(with $\alpha$ being an arbitrary parameter) does not change the monopole action~\eq{eq:S:mon:formal} due to
conservation condition~\eq{eq:closeness}. Therefore, in order to relate the theoretical and numerical results
we need to get the transverse part of the operator~\eq{eq:cS:expansion}.

\begin{figure}[ht]
\includegraphics[angle=-0,scale=0.75,clip=true]{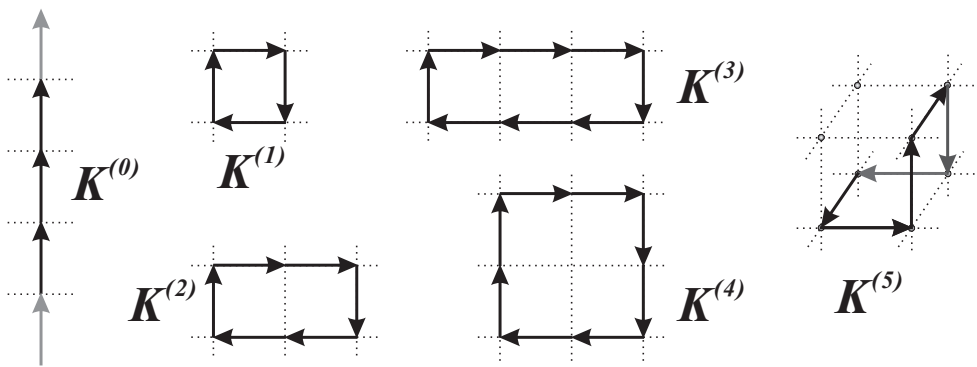}
\caption{Set of lattice currents used to get transverse elements $f_i$ of the monopole action operator.}
\label{fig:closed:curves}
\end{figure}

A simplest and, on the other hand, a practical way to do extract the transverse part of the quadratic
monopole operator is to calculate the monopole action on a set of closed monopole trajectories, $K^{(i)}$.
We consider six types of such monopole trajectories which are depicted in Figure~\ref{fig:closed:curves}.

Let us consider the analytical prediction for the transverse part of the monopole action.
Since we are working in the $\mu \gg q$ limit, we disregard $O(\mu^{-1})$ corrections to the quadratic
action\eq{eq:cS:expansion}. The validity of such approximation is discussed below.
The leading contribution to the monopole action evaluated on closed trajectories $K^{(i)}$ is
\beqn
f_i(b) \equiv \frac{S(K^{(i)})}{|K^{(i)}|} = \frac{2 \pi \, d_i}{\eta^2 b^2\, \Gamma(0, b^2 M^2_B t_{UV})}\,,
\label{eq:fi:leading}
\eeqn
where $|K^{(i)}|$ is the length of the trajectory $K^{(i)}$ and
\beqn
\begin{array}{rclcl}
d_0 & = & \cD^{(3)} (0,0,0) & = & 0.248028\dots\,,
\\
d_1 & = & \cD^{(3)} (0,0,0) - \cD^{(3)} (1,0,0)  & = & 0.166665\dots\,,
\\
d_2 & = & \cD^{(3)} (0,0,0) - \frac{2}{3}  \cD^{(3)} (1,0,0) - \frac{1}{3} \cD^{(3)} (2,0,0)  & = &  0.181055\dots\,,
\\
d_3 & = & \cD^{(3)} (0,0,0) - \frac{3}{4} \cD^{(3)} (1,0,0) - \frac{1}{4} \cD^{(3)} (3,0,0)  & = & 0.181292\dots\,,
\\
d_4 & = & \cD^{(3)} (0,0,0) - \cD^{(3)} (2,0,0) & = & 0.209836\dots\,,
\\
d_5 & = & \cD^{(3)} (0,0,0) - \frac{2}{3} \cD^{(3)} (1,0,0) - \frac{1}{3} \cD^{(3)} (1,1,0) & = & 0.176956\dots\,.
\end{array}
\label{eq:d0}
\eeqn
are the linear combinations of the values of the inverse three-dimensional Laplacian $\cD^{(3)}$
at certain points. The numerical values shown in Eq.~\eq{eq:d0} correspond to the lattice $48^3$.
Below we call the combinations $f^{(i)}$ of the couplings as "transverse couplings".

Using Table~\ref{tbl:gs} one can get the transverse combinations of couplings corresponding to the
numerically calculated action:
\beqn
\begin{array}{rclrcl}
f_0 & = & g_1 + 2 g_2 + 2 g_6 + 2 g_{22}\,,\quad & f_1 & = & g_1 - g_3\,,\\
f_2 & = & g_1 + \frac{2}{3} (g_2 - g_3 - g_4) - \frac{1}{3} g_7\,,\quad
& f_3 & = & g_1 + g_2 - \frac{3}{4} g_3 - g_4 + \frac{1}{2} g_6 - \frac{1}{2} g_{11} - \frac{1}{4} g_{23}\,,\\
f_4 & = & g_1 + g_2 - g_7 - g_{12}\,, \quad & f_5 & = & g_1 - \frac{2}{3} g_3 - \frac{1}{3} g_5\,.
\end{array}
\label{eq:fi:numerical}
\eeqn

Note that the transverse components of the analytical action~\eq{eq:fi:leading} with two free parameters
should describe six transverse combinations~\eq{eq:fi:numerical} obtained numerically.
We fit the $f_i$ components by~\eq{eq:fi:leading} independently for each $i=0,1,\dots,5$ and then compare
in Table~\ref{tbl:fitting:parameters} the fitting parameters $\eta$ and $m_{UV}$ as a self--consistency test.
Since we are working in the $\mu \gg 1$ limit we fitted the numerical data for the $n=6$ blocked monopoles.
A lower value of $n$ corresponds to the smaller scale $b$ and in this case we notice sizable deviations
of the numerical results from our fitting function. This is expected because we are working in the limit
$b \to \infty$. One the other hand, the higher value, $n=8$, correspond to the small lattice size of the coarse
lattice, $(N/n)^4=6^4$, which may lead to large finite--volume artifacts. Therefore we concentrated on
$n=6$ blocked monopoles.
\begin{table}
\begin{tabular}{|c|c|c|c|c|}
\hline
coupling & \multicolumn{2}{|c|}{$\eta \slash \sqrt{\sigma}$} &
           \multicolumn{2}{|c|}{$\sqrt{t}\, M_B \slash \sqrt{\sigma}$} \\
\hline
 & all clusters & max cluster &  all clusters & max cluster \\
\hline
$f_0$ &  0.521(25) &  0.509(23)  & 0.046(9) & 0.042(8)   \\
$f_1$ &  0.577(41) &  0.580(45)  & 0.020(9) & 0.022(10)  \\
$f_2$ &  0.565(34) &  0.537(37)  & 0.031(9) & 0.025(9)   \\
$f_3$ &  0.544(32) &  0.522(35)  & 0.032(9) & 0.026(9)   \\
$f_4$ &  0.554(28) &  0.532(30)  & 0.041(9) & 0.034(9)   \\
$f_5$ &  0.591(38) &  0.590(42)  & 0.025(9) & 0.026(10)  \\
\hline
{\bf average}  &  {\bf 0.552(13)}  & {\bf 0.534(13)} & {\bf 0.036(4)} & {\bf 0.031(4)}   \\
\hline
\end{tabular}
\caption{The values of the condensate $\eta$ and the ultraviolet cutoff $t_{UV}$ obtained in
a set of independent fits of the $n=6$ transverse monopole couplings~\eq{eq:fi:numerical}
by function~\eq{eq:fi:leading} for the all monopole clusters and for the percolating monopole cluster.
The best parameters of the overall fit of $f_1 \dots f_6$ are shown in the last row.}
\label{tbl:fitting:parameters}
\end{table}

\begin{figure}[htb]
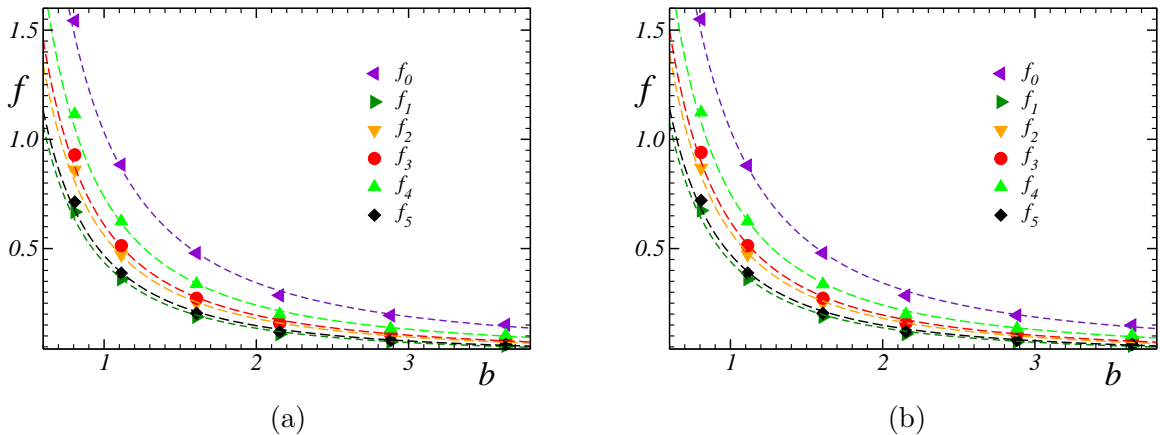

\begin{tabular}{cc}
\includegraphics[angle=-00,scale=0.3,clip=true]{couplings.all.eps}
&
\hskip 12mm\includegraphics[angle=-0,scale=0.3,clip=true]{couplings.max.eps}
\\
\hskip 5mm (a) & \hskip 15mm (b)
\end{tabular}
\caption{The fits of the transverse $n=6$ monopole couplings~\eq{eq:fi:numerical} by
function~\eq{eq:fi:leading} (a) for the all monopole clusters and (b) for the percolating monopole cluster.}
\label{fig:alpha:beta}
\end{figure}

The fits of the transverse couplings of the monopole action corresponding both to
all monopole clusters and to the percolating cluster are visualized in
Figures~\ref{fig:alpha:beta}(a) and (b), respectively.
The best fit parameters obtained from the fits of different transverse couplings $f_i$
(Table~\ref{tbl:fitting:parameters}) are very close to each other what provides a nice
self--consistency test of our approach. Moreover, the value of the monopole condensate $\eta$
calculated in large--$b$ limit from the all-clusters and the percolating cluster monopole action
are the same within error bars, as expected. The numerical value of the monopole condensate (obtained
by averaging of the results of the six independent fits) is $\eta = 243(6)$~MeV.

Finally, let us discuss the validity of the large $\mu$ approximation used in this paper. We are working
in the range of momenta $b \sqrt{\sigma} \sim 1 \dots 4$. The mass of the dual gauge boson obtained from
the fitting of the string profile by a classical string solution~\cite{string:profile} is
estimated as $M_B \approx 1.3~\mbox{GeV} \approx 3 \sqrt{\sigma}$. Therefore the value of $\mu$, Eq.~\eq{eq:mu},
is in the range $\mu \sim 3 \dots 12$. There are two types of corrections to our analytical results:
(i) the exponentially suppressed corrections to the operator $\cF^{-1}$ (discussed in Appendix~\eq{app:cF})
are smaller than 5\%; (ii) the $O(\mu^{-1})$ correction of Eq.~\eq{eq:cS:expansion} is of the order of
$10\%$ because of the local nature of the $\cG$ and $\cH$ operators, and due to low values of the
inverse Laplacian, $\cD^{(3)}(0,0,0) \approx 1/4$. Thus we estimate the systematic corrections to the value of
the monopole condensate to be of the order of $15\%$. Taking into account the systematic errors
we get $\eta = 243(42)$~MeV.

\section{Discussion and Conclusion}

We have obtained the value of the monopole condensate using the method of blocking from continuum to
the lattice. Namely, we have obtained numerically the effective monopole action in the Maximal
Abelian projection of quenched SU(2) lattice QCD. Then we have calculated analytically the
effective lattice monopole action starting from the continuum dual Ginzburg--Landau model. In our
simulations we restricted ourselves to the large values of the parameter $b$. This parameter
defines a scale at which the monopole charge is measured on the lattice.
In large--$b$ limit the action of the monopoles is dominated by the quadratic part, and higher monopole
interactions are suppressed. Thus in our analytical calculations we have neglected the quantum contributions
of the scalar monopole fields which are responsible for the higher order corrections to the
effective monopole lagrangian~\cite{chernodub}.

The comparison of the numerical and analytical results for the blocked action gives us the value of the
monopole condensate, $\eta = 243(42)$~MeV. This value is in a quantitative agreement with
another estimation of the monopole condensate, $\eta=194(19)$~MeV, obtained in Ref.~\cite{string:profile}
using a completely different method. Moreover, we have shown that our method is self--consistent, since is allows to describe various quadratic interaction of the monopole action
using approximately the same values of
the monopole condensate.

A few words about the ultraviolet cutoff $t_{UV}$ are now in order. Thus cutoff -- which enters the effective
monopole action~\eq{eq:cS:expansion} -- is an independent fitting parameter of the effective monopole action
at large scales, Eq.~\eq{eq:fi:leading}. In this paper we have neglected the fluctuations of the monopole scalar
fields since we were working at large scales $b$. Effectively, this corresponds to taking the London limit
of the Ginzburg--Landau model. The London limit possesses known logarithmic divergences ($i.e.$, the tension of the
Abrikosov vortex is logarithmically divergent function of an ultraviolet scale). The physics
of the monopole field fluctuations is "hidden" in the value of this cutoff. Strictly speaking, we have
to renormalize the model and consider the monopole field fluctuations to relate a logarithmic divergence
to the values of the physical parameters entering the lagrangian of the model.
This procedure becomes meaningful at small scales $b$.

At small values of the scale $b$ the higher order interactions (
four-point, six-point, $etc$) become
essential~\cite{chernodub}. Thus at short distances the scalar monopole field contribute
to the effective monopole action. From the point of view of the blocking from continuum,
at small values of $b$ the couplings of the monopole action become dependent on the parameters
of the potential of the monopole field. Thus, a comparison of the effective monopole action with
the blocked action at small scales $b$ may allow us to determine the shape of the monopole potential.
We discuss this problem in a forthcoming publication~\cite{in:preparation}.

\begin{acknowledgments}
Authors are grateful to F.V.Gubarev for useful discussions.
This work is supported by the Supercomputer Project of the Institute of Physical
and Chemical Research (RIKEN).
M.Ch. acknowledges the support by JSPS Fellowship No. P01023.
T.S. is partially supported by JSPS Grant-in-Aid for Scientific Research on
Priority Areas No.13135210 and (B) No.15340073.
\end{acknowledgments}


\appendix
\section{Proof of closeness of lattice monopole currents}
\label{app:closeness}

In order to prove the relation \eq{eq:closeness}
it is convenient to represent the lattice monopole current~\eq{eq:Link4D}
as the integral over momentum.  Using Eq.~\eq{eq:epsilon} and Eq.~\eq{eq:V:p} we get:
\beqn
\frac{1}{2} \epsilon_{\mu \nu \alpha \beta} \, \Sigma_{\alpha \beta}(C_{\gamma,s},x) =
i (p_\mu\, \delta_{\nu\gamma} - p_\nu\, \delta_{\mu\gamma})
\, b^3\, Q_\gamma(p b) \, e^{- i b (p,s)}\,,
\label{eq:epsilon:app}
\eeqn
where $\Sigma_{\alpha \beta}(C,x) \equiv \Sigma^{\partial C}_{\alpha \beta}(x)$,
the vector $Q_\alpha$ is given in Eq.~\eq{eq:Q}, and no summation
over index $\gamma$ is assumed. Then Eq.~\eq{eq:Link4D} can be rewritten as follows:
\beqn
 K_{s,\gamma} & = & - b^3 \int \!
 \frac{\dd^4 p}{(2 \pi)^4} \, (p_\mu\, \delta_{\nu\gamma} - p_\nu\, \delta_{\mu\gamma})\,
 \tilde{k}_\mu(-p)\, \frac{p_\nu}{p^2}\,  Q_\gamma(p b) \, e^{- i b (p,s)} \nonumber \\
 & = &  - b^3 \int \!
 \frac{\dd^4 p}{(2 \pi)^4} \,
 \tilde{k}_\gamma(-p)\,  Q_\gamma(p b) \, e^{- i b (p,s)}\,,
\label{eq:K:p}
\eeqn
where $\tilde{k}_\mu(p) = \int \dd x \, k_\mu(x)\, e^{- i (p,x)} $ is the Fourier transformed
continuum monopole current. There is no summation
over index $\gamma$ in Eq.~\eq{eq:K:p}.  To get the second line of Eq.~\eq{eq:K:p} we used
the closeness condition of the continuum monopole currents,
\beqn
p_\mu \tilde{k}_\mu(p) = 0\,.
\label{eq:dp:app}
\eeqn

According to Eq.~\eq{eq:C} the lattice monopole currents, $K_{s,\mu}$, are associated with the
centers of the three--dimensional cubes $C_{s,\mu}$. The positions of the cube centers
are characterized by the integer--valued coordinates $s$. The corners of the cubes belong
to the original lattice while the monopole currents themselves are associated with the dual lattice.
The sites of the dual lattice are shifted by the four--dimensional vector $(1/2,1/2,1/2,1/2)$
with respect to the sites of the original lattice. Thus, the center of the cube $K_{s,\mu}$
does not belong to the dual lattice because the $s_\mu$ coordinate of the center of the cube
corresponds to the time--slice of the original lattice. In our coordinates,
the monopole current defined on the cube $K_{s,\mu}$ must be associated with the point
$\dual s = s + \hat \mu/2$ belonging to the dual lattice.

Thus, the closeness condition~\eq{eq:closeness} at the site $\dual s$ of the dual lattice is
\beqn
(\partial' K)_{\dual s} & = & \sum^4_{\gamma=1} \Bigl(K_{\dual s,\gamma} -
K_{\dual s - \hat \gamma,\gamma}\Bigl) \equiv
\sum^4_{\gamma=1} \Bigl(K_{s + \frac{1}{2} \hat \gamma,\gamma}
- K_{s - \frac{1}{2} \hat \gamma,\gamma}\Bigl) \nonumber \\
& = & 2i b^3 \int \! \frac{\dd^4 p}{(2 \pi)^4} \,
 \tilde{k}_\gamma(-p)\, \sin (b p_\gamma/2) \, Q_\gamma(p b) \, e^{- i b (p,s)}\,.
\label{eq:closeness:app1}
\eeqn
Using Eq.~\eq{eq:Q} we notice that $2 \sin (b p_\gamma/2) \, Q_\gamma(p b) = p_\gamma \, Q(p b)$,
where the quantity $Q(x) = \prod\nolimits_{\nu} \frac{\sin x_\nu /2}{x_\nu/2}$ does not carry any
Lorentz index. Then Eq.~\eq{eq:closeness:app1}
together with the conservation of the continuum monopole charge~\eq{eq:dp:app}
implies the closeness of the lattice monopole currents,
\beqn
(\partial' K)_s = i b^3 \int \! \frac{\dd^4 p}{(2 \pi)^4} \, p_\gamma
 \tilde{k}_\gamma(-p)\,  \, Q(p b) \, e^{- i b (p,s)} \equiv 0\,. \nonumber
\eeqn

\section{Calculation of the operator ${\mathbf{\cF^{-1}}}$}
\label{app:cF}

In this Appendix we calculate the expression for the inverse operator $\cF^{-1}$,
presented in Eq.~\eq{eq:F:inverse}, for $\mu \equiv M_B b \gg 1$.

Let us consider first the diagonal components of the inverse operator $\cF^{-1}$. Without
loss of generality we take $\mu=\nu=4$ and $s'=0$. We get:
\beqn
\cF^{-1}_{s0,44} = \int \frac{\dd^4 q}{(2 \pi)^4} \, \frac{{\vec q}^2}{q^2_4 + {\vec q}^2 + \mu^2} \,
\prod^3_{i=1} {\Bigl(\frac{\sin q_i/2}{q_i/2}\Bigr)}^2 \cdot e^{i q_4 s_4 + i (\vec q, \vec s)}\,.
\label{eq:app:F44}
\eeqn
It is convenient to introduce the additional integral,
\beqn
\frac{1}{q^2_4 + {\vec q}^2 + \mu^2} = \int^\infty_0 \dd t \, e^{-(q^2_4 + {\vec q}^2 + \mu^2) t}\,,
\eeqn
and represent the integral~\eq{eq:app:F44} in the form:
\beqn
\cF^{-1}_{s0,44} = \int\limits^\infty_0 \dd t \, e^{- \mu^2 t }\, P_0(s_4, t) \,
\sum_{i=1}^3  P_1(s_i,t) \prod_{\stackrel{j=1}{j \neq i}}^3 P_2(s_j,t)\,,
\label{eq:app:F44:P}
\eeqn
where
\beqn
P_0(s,t) & = & \int\limits_{-\infty}^\infty \frac{\dd q}{2 \pi}\, e^{- t q^2 + i q s} =
\frac{1}{2 \sqrt{\pi t}} \, e^{- s^2/ 4 t}\,,\\
\label{eq:P0}
P_1(s,t)  & = & 4 \!\! \int\limits_{-\infty}^\infty  \!\! \frac{\dd q}{2 \pi}\, \sin^2 \frac{q}{2}\cdot
e^{- t q^2 + i q s}
= - \frac{1}{2 \sqrt{\pi t}} \Bigl(e^{-(s+1)^2/4t} + e^{-(s-1)^2/4t} - 2 e^{-s^2/4t}\Bigr)\,,
\label{eq:P1}\\
P_2(s,t)  & = & \int\limits_{-\infty}^\infty \frac{\dd q}{2 \pi}\, {\Bigl(\frac{\sin q/2}{q/2}\Bigr)}^2\,
e^{- t q^2 + i q s} =\sqrt{\frac{t}{\pi}} \Bigl(e^{-(s+1)^2/4t} + e^{-(s-1)^2/4t} - 2 e^{-s^2/4t}\Bigr)
\nonumber\\
& & + \frac{1}{2} \Bigl[(s+1) \Erf\Bigl(\frac{s+1}{2 \sqrt{t}}\Bigr) +
(s-1) \Erf\Bigl(\frac{s-1}{2 \sqrt{t}}\Bigr) - 2 s \Erf\Bigl(\frac{s}{2 \sqrt{t}} \Bigr) \Bigr]\,,
\label{eq:P2}
\eeqn
and $\Erf(x)$ is the error function,
\beqn
\Erf(x) = \frac{2}{\sqrt{\pi}} \int^x_0 \dd y \, e^{- y^2}\,.
\nonumber
\eeqn

To calculate the off-diagonal components of the inverse operator $\cF^{-1}$ we take
$(\mu,\nu)=(1,2)$ and $s'=0$ (again, without any loss of generality):
\beqn
\cF^{-1}_{s0,12} & = & \int \frac{\dd^4 q}{(2 \pi)^4} \, \frac{- q_1 q_2}{q^2 + \mu^2} \,
\prod^2_{i=1}\frac{\sin q_i/2}{q_i/2} \cdot
\prod^4_{j=3}{\Bigl(\frac{\sin q_j/2}{q_j/2}\Bigr)}^2 \cdot e^{i (q, s)} \nonumber\\
& = & - \int_0^\infty \dd t \, e^{- \mu^2 t} \,
\prod^2_{i=1} P_3(s_i,t)
\cdot \prod^4_{j=3} P_1(s_i,t)
\label{eq:app:F12}
\eeqn
where
\beqn
P_3(s,t) & = & 2 \int\limits_{-\infty}^\infty \frac{\dd q}{2 \pi}\, \sin \frac{q}{2} \cdot
e^{- t q^2 + i q s} =
\frac{i}{2 \sqrt{\pi t}} \Bigl(e^{-(2s-1)^2/16t} - e^{-(2s+1)^2/16t}\Bigr)\,.
\label{eq:P3}
\eeqn

Eqs.(\ref{eq:app:F44}-\ref{eq:P3}) represent the exact expressions for the diagonal and off-diagonal
elements of the inverse operator $\cF^{-1}$. Unfortunately, due to the presence of the $\Erf$--functions
in $P_1$, Eq.~\eq{eq:P1}, the integrals \eq{eq:app:F44} and \eq{eq:app:F12}
can not be taken analytically. However, in the limit $\mu \to \infty$, which corresponds according to
Eq.~\eq{eq:mu} to large blocking scales $b$, leading contributions to these integrals can be easily
estimated.

Let us first consider Eq.~\eq{eq:app:F44}. The main contribution to this integral is coming from the
region of small $t$. At small $t$ the error function can be represented as
\beqn
\Erf(x) = \sign(x) - \frac{e^{-x^2}}{\sqrt{\pi} x}  (1 + O(x^{-2}))
\quad \mbox{for} \quad x\gg 1\,.
\label{eq:Erf}
\eeqn
Therefore at general values of $s$ the expression~\eq{eq:app:F44} is given by a sum of integrals
of the form
\beqn
I(\mu, \tilde s) = \int_0^\infty \dd t \, \exp\{ - \mu^2 t - \tilde s^2/t + C\, \log t\}\,,
\label{eq:I}
\eeqn
where $C$ is a constant of the order of unity and the quantity $\tilde s$ depends on the value of $s$
({\it i.e.}, $\tilde s = s/2, (s-1)/2$, {\it etc.}).
The value of $\tilde s$ is either of the order of unity or zero. At $\tilde s \sim 1$ and large $\mu$
we get $I(\mu, \tilde s) \sim \exp\{- 2 \mu \tilde s\} \ll 1$. Thus the integral~\eq{eq:I} with $\tilde s \neq 0$
is exponentially suppressed and therefore it will be neglected below. The leading contribution
to the operator $\cF^{-1}$ comes from the integrals of the form~\eq{eq:I} with $\tilde s = 0$ which
are saturated at small $t$.

Using the expansion~\eq{eq:Erf} we get in the leading order in the limit $t \to 0$:
\beqn
P_0(s,t) & = & \frac{1}{2 \sqrt{\pi t}} \, \delta_s + O(e^{-\const /t})\,,
\label{eq:P0:t}\\
P_1(s,t) & = & - \frac{1}{2 \sqrt{\pi t}} \, \Delta_s + O(e^{-\const /t})\,,
\label{eq:P1:t}\\
P_2(s,t) & = & \sqrt{\frac{t}{\pi}} \, \Delta_s + \delta_s + O(e^{-\const /t})\,,
\label{eq:P2:t}\\
P_3(s,t) & = & O(e^{-\const /t})\,,
\label{eq:P3:t}
\eeqn
where
\beqn
\delta_s =
\left\{
\begin{array}{ll}
1, & s = 0 \\
0, & \mbox{otherwise}
\end{array}
\right.\,, \qquad
\Delta_s =
\left\{
\begin{array}{rl}
1,   & s = 1,-1 \\
- 2, & s = 0 \\
0, & \mbox{otherwise}
\end{array}
\right.\,,
\label{eq:laplacian}
\eeqn
are the Kronecker symbol and the one-dimensional lattice Laplacian, respectively.

According to Eqs.~(\ref{eq:app:F12},\ref{eq:P3:t}),
 the elements with $\mu\neq\nu$ of the operator
$\cF^{-1}$ are exponentially suppressed, $\cF^{-1}_{\mu\neq\nu} \sim O(e^{-\const \mu})$. As for the
diagonal elements of this operator, $\mu=\nu$, we get
\beqn
\cF^{-1}_{s0,44} & = & \frac{1}{4 \pi^2} \int\limits_0^\infty \dd t\, e^{-\mu^2 t} \, \delta_{s_4}
\sum_{i=1}^3 \Delta_{s_i} \, \prod_{\stackrel{j=1}{j\neq i}}^3 \Bigl(\Delta_{s_j} +
\sqrt{\frac{\pi}{t}} \delta_{s_j}\Bigr) + O(e^{-\const \mu}) \label{eq:F44:calc2}\\
& = & \frac{1}{4 \pi} \delta_{s_4} \, \Bigl[
\Gamma(0, t_{UV} \mu^2) \, \Bigl(\Delta_{s_1} \delta_{s_2} \delta_{s_3} + \mbox{cyclic}\Bigr)
+ \frac{2}{\mu} \, \Bigl(\Delta_{s_1} \Delta_{s_2} \delta_{s_3} + \mbox{cyclic}\Bigr) \nonumber\\
& & + \frac{3 }{\pi \mu^2} \Delta_{s_1} \Delta_{s_2} \Delta_{s_3} \Bigr] + O(e^{-\const \mu})\,.
\nonumber
\eeqn
Here $\Gamma$ is the incomplete gamma function, $\Gamma(a,x) = \int_x^\infty t^{a-1}\, e^{-t} \, \dd t$,
and "cyclic" means cyclic permutations over the indices $s_i$. To get Eq.~\eq{eq:F44:calc2}
we used Eqs.~(\ref{eq:app:F44:P},\ref{eq:P0:t},\ref{eq:P1:t},\ref{eq:P2:t}).
We also introduced the ultraviolet cutoff,
$t_{UV}$, to regularize the logarithmically divergent piece of Eq.~\eq{eq:F44:calc2}. Noticing that
$\cD^{(3)}_4 (\vec s) = \Delta_{s_1} \delta_{s_2} \delta_{s_3} + \mbox{cyclic}$ is the three
dimensional Laplacian we get the final expression for $\cF^{-1}$ presented in Eq.~\eq{eq:cF:inverse:final}.

\end{document}